Università degli studi di Salerno
Dipartimento di informatica

Università degli studi
di Salerno

# Cryptanalysis of the SLAP Authentication Protocol

Tesi di Laurea Triennale
in
Informatica

*Supervisor*:
Prof. Roberto De Prisco

*Student*:
Giovanni Ciampi
Matr. 0512102999

Anno Accademico 2016/2017

Premise:
This is my bachelor thesis work, carried out at the University of Salerno between February and September 2017, under the supervision of Prof. Roberto De Prisco, and with the collaboration of Prof. Paolo D'Arco and the student Domenico Desiato. The final work led to the publications "Confusion and Diffusion in Recent Ultralightweight RFID Authentication Protocols", that can be found at https://cryptacus.cs.ru.nl/slides/booklet.pdf and to "Design Weaknesses in Recent Ultralightweight RFID Authentication Protocols", that can be found at https://link.springer.com/chapter/10.1007/978-3-319-99828-2_1 .

# Contents

# Abstract

RFID (Radio Frequency Identification) is a powerful technology that, due to its numerous advantages, is supposed to replace the various identification systems such as barcodes or magnetic stripes in a short time. There are three devices involved in an RFID protocol : a Reader, a Tag, and a back-end Database. One of the key factors for the RFID technology is that, in order to be used on large scale, the price of the Tags has to be cheap: it cannot be expensive because who is supposed to use it would need a great amount of them, furthermore, Tags must have very small dimensions. The low-cost nature of such devices implies that it is impossible to use traditional cryptographic protocols on them, furthermore, there are certain Tags that could not even generate random numbers or use basic hashing functions, these Tags are called Ultra-Lightweight. Many experts are trying to build secure protocols that involve only simple bitwise logical operations for the Ultra-Lightweight Tags, but, unfortunately, each one of these protocols, that was supposed to be secure, turned out to be vulnerable to some serious attack after short time from pubblication. The need for a secure RFID authentication protocol today seems more urgent than ever, because this technology is already used for security purposes, such as electronic passports. The challenge to build a secure protocol for Ultra-Lightweight RFID systems is so hard that in several years no one has been able to build one. In this thesis we analyse in great detail a recently proposed protocol for Ultra-Lightweight RFID systems called SLAP [1], with the aim of finding new vulnerabilities. SLAP has already been violated (along with a set of similar protocols) by *Safkhani* and *Bagheri* [5], that have recently published a de-synchronization attack. At the end of our analysis, we will propose an impersonification attack to the protocol, along with a fix for our attack and some considerations on the attacks proposed on this kind of protocols.

# Chapter 1

# Introduction

The RFID (acronym of Radio-Frequency Identification) is a technology which allows automatic recognition of things. The RFID schema consists of three entities: a Reader, a Tag, and a back-end Database. Through radiofrequencies, the Reader can communicate with the Tag and read/update its contents. This technology can be used in a variety of fields: for example electronic passports, credit cards, warehouse logistics, etc; and, in the last years, it is rapidly spreading thanks to its numerous advantages over the others common identification systems (like barcodes or magnetic stripe-systems). The advantages over the traditional identification systems are countless, for example, compared to a barcode, an RFID Tag doesn't need to be visible to be read, and it is possible to read many Tags at the same time. Moreover, the Tags are waterproof and heat-resistant. To get a practical idea of how these advantages are valuable, just think that in a certain warehouse, if the items were identified by RFID Tags, in a single instant it would be possible to know all the contents of a box full of items, without opening it. The functioning mechanism is very simple: the Tag contains a series of informations regarding, for example, an item; the Reader is capable to read/uptate that information, and the back-end Database manages all that kind of information. For this technology to have sense, each entity of the system must satisfy certain requirements, in particular, there are certain important restrictions concerning the building of the Tag: size and price.

Indeed, for this autentication system to be effective, we don't want the size of the Tags to be relevant: instead, we want them being so small that we don't need to care about (how could you insert a few centimeters-big Tag inside a credit card?). Furthermore, we don't even want to care about their price: even a small cost, multiplied by the number of thousands of items in a warehouse, could make their use less desirable. In the real world, Tags are so small that they can be easily put inside credit cards, and their prices is, in the most of the cases, below 0.05$. Because of the important restrictions on size and price, it is normal that even performances are limited: in most cases the total amount of logic gates on a Tag is between 5'000 and 10'000, and of these logic gates only a small amount (between 250 and 3'000) can be used for security purposes. You may wonder why the security of the RFID systems is a questions that matters, thinking about RFID Tags used to store some informations on an item inside a clothes shop: firstly, it is important to stress the fact that the Readers not only can *read*, but they can also *write* inside a Tag, and we don't want a potential *Attacker* to be capable of randomly update prices inside a shop; secondly, these tiny devices are also used to store important information about, for example, passports, credit cards, and even dogs! In this section, we introduced why security of Reader-Tag communication is important, without mentioning the last entity of our system, the Database. Since the Database and the Reader don't have important restrictions, they can secure their communications through traditional cryptographic protocols, and so we don't need to care about that.

## 1.1 RFID Authentication Protocols

Because of the impossibility of using traditional cryptographic protocols in the communication between Reader and Tag, and since there isn't any known secure protocol that can work with a so little computing power, many new protocols are being devised for this specific purpose. These protocols are divided into four categories, organized according to the power of the devices:

1. **Full-Fledged**: These protocols are suitable for very powerful Tags,

which have enough computing power that they can incorporate traditional cryptography, like one-way hashing functions.

2. **Simple**: This kind of protocols support pseudo-random number generation and one-way hashing functions only.

3. **Lightweight**: This cathegory of protocols refers to Tags being capable of generating random numbers, and computing simple functions like CRC (Cyclic Redundancy Code) ckecksum, but not hashing functions.

4. **Ultra-Lightweight**: These protocols are referred to the majority of Tags, that are the Tags that can only compute simple bitwise logical operations, like AND, OR, XOR, etc. Random number generation isn't possible by this kind of Tag.

The most interesting category, as regards research, is the category of *Ultra-Lightweight* protocols: it may seem impossible to keep important data like passports informations, or credit cards codes, secure, through a protocol that uses only simple bitwise logical operations... and indeed someone thinks it is. This challenge has involved many experts, and for this reason every year many new *Ultra-Lightweight* protocols are proposed. Even though many new protocols arised, there is still not a single *Ultra-Lightweight* protocol that is known to be secure, on the contrary, citing the authors of the SLAP protocol: *"Unfortunately, it is quite usual that the time for publication of a serious attack on a new scheme is extremely short. This is a good phenomenon and at the same time gives us some admonishment".*

## 1.2 The *GUMAP* Protocols

A series of recently proposed Ultra-Lightweight protocols (between 2015 and 2017) has got a structure very similar to a first protocol called $RAPP$. Some examples are the protocols $R^2APP$, $RCIA$ [3], $KMAP$ [4], $SLAP$ [1] and $SASI+$ [2]; they are identified under the name $GUMAP$ (Generalized Ultra-Lightweight Mutual Autentication Protocols). These protocols follow a similar structure: first of all, both the Tag and the Reader know two $IDs$:

$ID_{old}$ and $ID_{new}$, and each $ID$ have an associated couple of keys: $k_1^{new}$ $k_2^{new}$, and $k_1^{old}$ $k_2^{old}$. The protocol starts with an *HELLO* message sent from the Reader to the Tag, that, once received the message, responds with its $ID^{new}$, which is used by the Reader to fetch the associated couple of keys, $k_1^{new}$ and $k_2^{new}$, from the Database. If the Reader isn't able to associate the received $ID$ with a couple of keys, it sends another $HELLO$ to the Tag, which then will respond with its $ID_{old}$. At this point, if the Reader has a couple of associated keys, the protocol will be executed with these keys and $ID$, otherwise, the protocol will be aborted.
Once it has fetched the keys, the Reader generates a random number $n$, and the three values ($k_1$, $k_2$ and $n$) are given in input to a certain function, that varies from one protocol to another, and the computed message is sent to the Tag, that, knowing $k_1$ and $k_2$, is able to retrieve the random number $n$. The aim of this operation is to hide the number $n$, in a way that let us sending it through an insecure channel, without the risk that an attacker who eavesdropped the message would identify it. At this point both the Tag and the Reader knows the three values, and they exchange some other messages to understand if the values on the other side are the same; then, if the Tag or the Reader notices that the values on the other device are different, the protocol is aborted, otherwise, a new triplet of values $ID^{new}$ $k_1^{new}$ $k_2^{new}$ is generated and the values are updated on both sides.

### 1.2.1 Some Practical Motivations for the *GUMAP* Protocols

There is a significant difference that distinguishes the $GUMAP$ protocols from $RAPP$: the fact that $RAPP$ has only one triplet $ID$, $k_1$ and $k_2$, and not two. This is an important point because it arose from a necessity, in fact soon after the pubblication of $RAPP$, a simple desynchronization attack was published (recall the quote from the $SLAP$ authors in *section 1.1*). Altough the protocol presented this great vulnerability, its global structure seemed to be interesting to the other authors, which decided to try to upgrade it without reinventing protocols from scratch. To overcome this vulnerability, the creators of the $GUMAP$ protocols decided to store each time not only the new triplet of generated values, but also the old one, adding the triplet:

$k_1^{old}$, $k_2^{old}$ and $ID_{old}$ to the set of stored values.
Let's see what are the effects of this change: we said that the vulnerability of *RAPP* allows us a desynchronization attack, that is, only the triplet of values on the Tag (or that on the Reader) is modified, and so the two devices cannot communicate anymore (the various kinds of attack will be described with more details in *section 2.3*). Let's suppose that we want to try to do the same attack on a *GUMAP* protocol, so we update the values stored on the Tag, but the values on the Reader stays the same. At this point, the Tag contains the triplets ($k_1^{new}$, $k_2^{new}$, $ID_{new}$), ($k_1^{new'}$, $k_2^{new'}$, $ID_{new'}$), while the Reader contains ($k_1^{old}$, $k_2^{old}$, $ID_{old}$), ($k_1^{new}$, $k_2^{new}$, $ID_{new}$). When they'll try to communicate, the Reader will send the *HELLO* message to the Tag, which will send back $ID_{new'}$, that is unknown to the Reader. At this point, the Reader will send another *HELLO* to the Tag, that, this time, will send its $ID_{new}$. After this, the Reader knows the *ID*, and so the communication can happen. This is the proof that they've not been desynchronized.

### 1.2.2 The SLAP Protocol

The *SLAP* protocol was proposed by *Luo*, *Su* and *Huang* in 2016, and it is one of the so called *GUMAP* protocols. The name *SLAP* stands for *S*uccint and *L*ightweight *A*uthentication *P*rotocol. Let's start by defining the operations used by the protocol: the first operation, very common in Ultra-Lightweight protocols, is the function $Rotation(a, b)$ (or simply $Rot(a, b)$). This function is really simple: it takes as input two strings and computes a circular left shift of the first, where the number of positions being shifted, corresponds to the Hamming Weight of the second. The *Rotation* function works in the following way: suppose that $a = a_{n-1}a_{n-2} \cdot \cdot a_1 a_0$ is a binary string and we want to compute $Rot(a, b)$: first of all, we compute $wt(b) = k$, then we divide $a$ into the two substrings $a_1 = a_{n-1} \cdot \cdot a_{n-k-1}$ and $a_2 = a_{n-k-2} \cdot \cdot a_0$, we have $Rot(a, b) = a_2 a_1$. Later in the text, we'll find very often the expression $Rot(a)$: it simply stands for $Rot(a, a)$, so the string $a$ is rotated by a number of positions equals to its own Hamming Weight.
The second function used during the protocol, is the function *Conversion(a, b)*. It has been defined by the authors of the protocol, and it's not so easy to explain, so we'll dedicate all the next section to it. For now, it's enough to know that this function takes as input two strings of the same length, and a fixed value $Treshold$; from this input, the function computes a third string (with the same length of $a$ and $b$) with the aim that the output string seems as much random as possible.
The following schema represents the phases of the protocol and shows how the various values are computed:

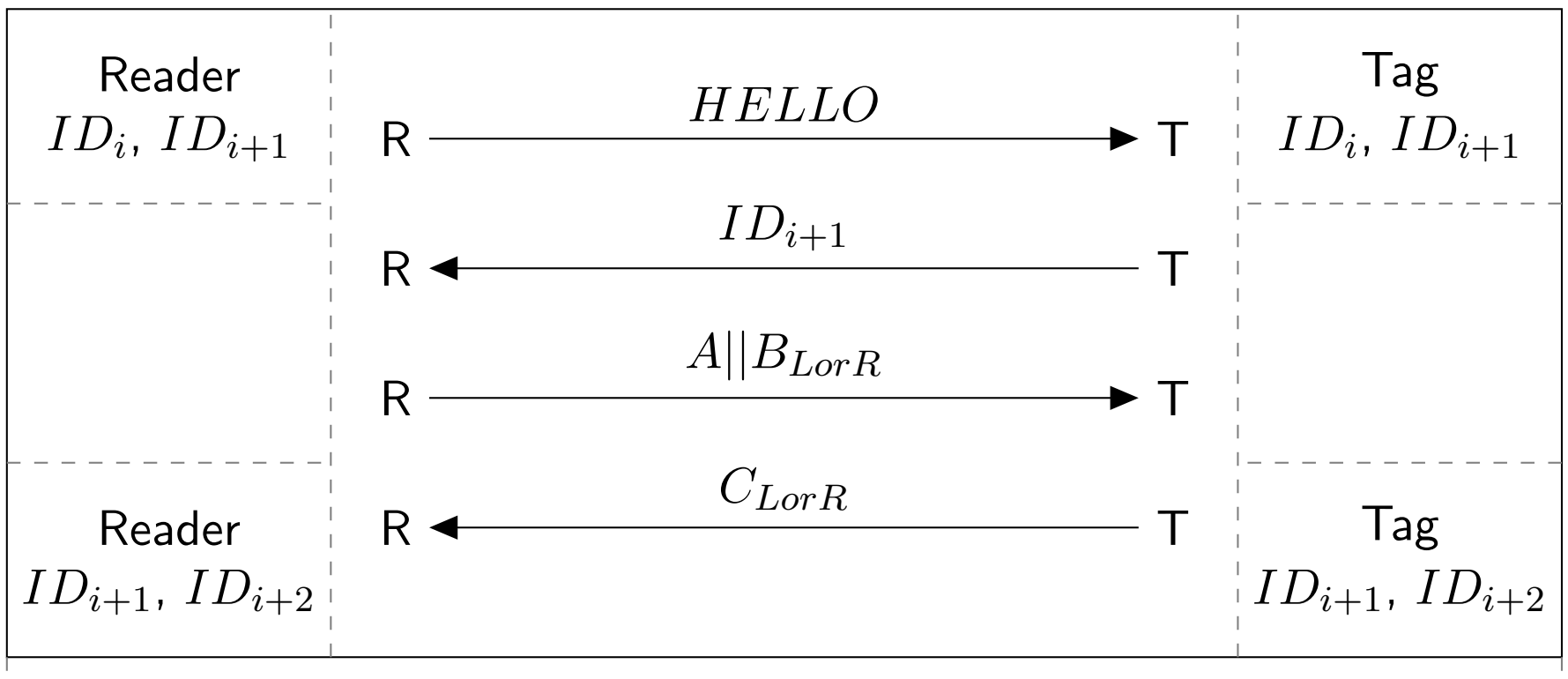

$$A = Con(k_1^{new}, k_2^{new}) \oplus n$$
$$B = Con(Rot(k_1^{new}, n), k_1^{new} \oplus k_2^{new}) \oplus Rot(Con(k_2^{new}, k_2^{new} \oplus n), k_1^{new})$$
$$C = Con(Con(B, k_1^{new}), Con(k_1^{new}, k_2^{new} \oplus n)) \oplus ID$$

$$k_1^{new} = Con(k_1, n) \oplus k_2$$
$$k_2^{new} = Con(k_2, B) \oplus k_1$$
$$ID^{new} = Con(ID, n \oplus (B''_{LorR} || C''_{LorR}))$$

Figure 1.1: The $SLAP$ Protocol

Now, let's start with the protocol: at first, like the other *GUMAP* protocols, both the Reader and the Tag know the values $ID^{new}$, $ID^{old}$, $k_1^{new}$, $k_2^{new}$, $k_1^{old}$ and $k_2^{old}$. These are the steps of the protocol:

1. The Reader sends an *HELLO* message to the Tag. This message doesn't contain any information.

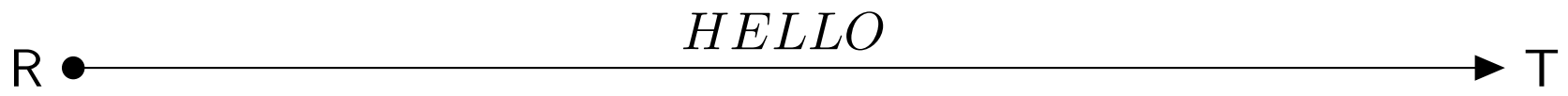

2. The Tag sends its $ID_{new}$ to the Reader.

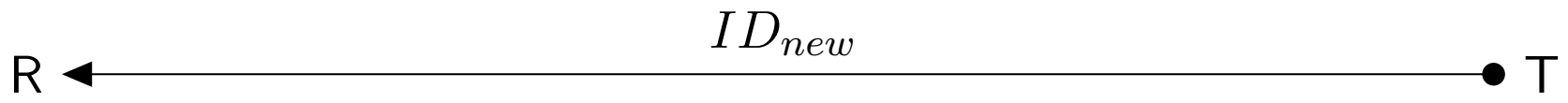

3. The Reader tries to fetch from the Database the keys associated with the recived $ID$: if it succeeds, it skips to the next point. If the Reader is not able to retrieve the keys the first time, it sends another *HELLO* to the Tag: this time the Tag will send its $ID_{old}$ to the Reader, and, if the Reader finds a match in the Database, the protocol continues, otherwise, it is aborted. From now on, we'll assume that the communication succeeds the first time, and so we assume that the devices use $ID_{new}$, $k_1^{new}$ and $k_2^{new}$: so the only difference with the other situation is the name of the variables used (to be more specific, during the second attempt, the devices try to use the values $ID_{old}$, $k_1^{old}$ and $k_2^{old}$).

4. The Reader generates a random number $n$, and it uses this three values to compute two strings: the first string is

$$A = Conv(k_1^{new}, k_2^{new}) \oplus n$$

and the second String is

$$B = Conv(Rot(k_1^{new}, n), k_1^{new} \oplus k_2^{new}) \oplus Rot(Conv(k_2^{new}, k_2^{new} \oplus n), k_1^{new})$$

At this point, the Reader sends to the Tag the messages $A$, and $B_R/B_L$ (where $B_R$ indicates the right half of $B$ and $B_L$ the left half), depending on if the Hamming Weight of $B$ is even or odd (if it's even it sends $B_R$ and vice-versa).

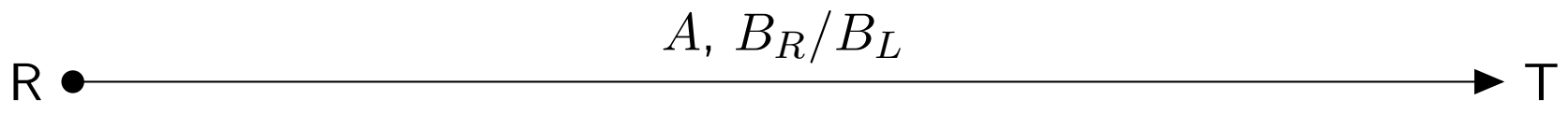

5. Once it has received $A$ and $B_R/B_L$, the Tag extracts $n$ from $A$, by computing $Conv(k_1^{new}, k_2^{new})$ and by XORing this value with $A$. Immediately afterwards, the Tag uses $n$ to compute $B$ and so $B_R/B_L$; then, it compares the value with the received one, and, if both the values are equals, then the Tag computes a series of new values ($ID$, $k_1$ and $k_2$) and updates its content by storing the values $ID^{new}$, $k_1^{new}$ and $k_2^{new}$ as $ID^{old}$, $k_1^{old}$ and $k_2^{old}$; and the newly computed values as $ID^{new}$, $k_1^{new}$ and $k_2^{new}$. If the value that the Tag received is different from the one that it calculated, then the protocol is aborted.

6. At this point, the values on the Tag are different from the values on the Reader; so it is necessary to update them in the same way. So the Tag computes a value

$$C = Con(Con(B, k_1^{new}), Con(k_1^{new}, k_2^{new} \oplus n)) \oplus ID$$

and it sends the half of $C$, $C_L/C_R$ with the same logic it has used for $B$.

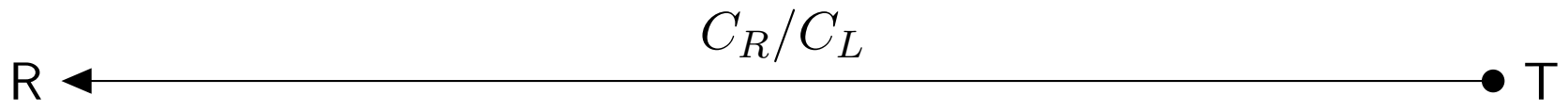

When it receives this last value, the Reader uses the same formula to compute its own $C$. Then, if the two values are equals, even the triplet on the Reader is updated, otherwise, the two stored triplets stays the ame as before.

### 1.2.3 The *Conversion* Function

One of the strengths of the SLAP protocol is the *Conversion* function. This function takes as input two strings of fixed length, a fixed value *Treshold*, and computes an output value whose aim is to seem as much random as possible.

Let's see how it works:
Let's suppose that the length of the two input strings is fixed to 32, and that the Treshold value is set to $T = 6$. *Conversion* takes as input two binary strings $A = a_{31}a_{30}\cdots a_1a_0$ and $B = b_{31}b_{30}\cdots b_1b_0$ (it is important to notice that the bits are numbered in decreasing order, from the left to the right), then, the function consists of three phases:

1. **Grouping**: In this phase, each string is recursively splitted into several substrings, each of which has to be smaller than $T$, depending on its Hamming Weight. Here is an example: suppose $A = a_{31}a_{30}\cdots a_1a_0$ and $wt(A) = 20$, then $A$ is splitted into $A_1$ and $A_2$ where $A_1 = a_{20}a_{19}\cdots a_1a_0$ and $A_2 = a_{31}a_{1930}\cdots a_{22}a_{21}$, and so $length(A_1) = wt(A)$ and $length(A_2) = length(A) - wt(A)$. This is done recursively for each substring until each of them has length smaller or equal than $T$. The same is done with $B$. The output of this phase is an ordered set of substrings for each one of the two input strings.

2. **Rearrange**: In this second phase, we use the divisions that we computed on one string to split the other, in other words, we exchange the divisions. Here is an example: suppose that the output of the previous phase is $A = (a_{31}\cdot\cdot a_{28})(a_{27}\cdot\cdot a_{23})(a_{22}\cdot\cdot a_{17})\cdot\cdot\cdot\cdot(a_3\cdot\cdot a_0)$, $B = (b_{31}\cdot\cdot b_{27})(b_{26}\cdot\cdot b_{25})(b_{24}\cdot\cdot b_{19})\cdot\cdot\cdot\cdot(b_4\cdot\cdot b_0)$, then we exchange this groups from $A$ to $B$ and from $B$ to $A$, in the following way: $A = (a_{31}\cdot\cdot a_{27})(a_{26}\cdot\cdot a_{25})(a_{24}\cdot\cdot a_{19})\cdot\cdot\cdot\cdot(a_4\cdot\cdot a_0)$, $B = (b_{31}\cdot\cdot b_{28})(b_{27}\cdot\cdot b_{23})(b_{22}\cdot\cdot b_{17})\cdot\cdot\cdot\cdot(b_3\cdot\cdot b_0)$. After this operation, we apply on each substring the function *Rotation* and we consider $A$ and $B$ as the concatenations of the corresponding lists of rotated substrings. Now we have this two strings $A$ and $B$, that are rearranged in a completely different way, compared to the beginning.

3. **Composition** This is the third and last phase and it simply consists in a XOR between the new $A$ and $B$ resulting from the previous phase (we can compute it since both the strings have equal length).

An explanation of this function in words may result not so clear, so, let's see how it works by a practical example:
Let's start by defining the input variables of the function:

$$A = 11101001010111100101011010011011$$

$$B = 10111001010100100101011011101100 0$$

$$T = 6$$

So it's clear that we are using 32 bit-long strings. The first phase is **Grouping**, so we start by calculating $wt(A)$, that is equal to 19. Now we split the string $A$ into the two substrings $A_1$ and $A_2$ so that $A_2 = a_{31} \cdot \cdot a_{19}$ and $A_1 = a_{18} \cdot \cdot a_0$ (during this phase, it is important to remember that the characters of the strings are numbered in decreasing order from the left to the right!). So, in this case, $A_1 = 1100101011010011011$ and $A_2 = 1110100101011$. At this point we have to measure the length of each substring, and, if it's greater than the treshold, then we need to split it again as we did before. As we can easily notice, $length(A_1) = 19 > T$, so we repeat the same thing that we did for $A$: $wt(A_1) = 11$, so $A_1 = A_{12}A_{11}$ such that $A_{12} = a_{18} \cdot \cdot a_{11}$ and $A_{12} = a_{10} \cdot \cdot a_0$, from that we get $A_{12} = 11001010$ and $A_{11} = 11010011011$. The same thing happens with $A_2$, because $length(A_2) = 13 > T$; once splitted, $A_2$ looks like that: $A_2 = A_{22}A_{21}$ where $A_{22} = 11101$ and $A_{21} = 00101011$. At this point, an interesting thing happened: we have $A_{22} = 11101$ such that $length(A_{22}) \leq T$, so we don't need to split it anymore.
Let's see where we are right now, we split $A$ in the following way:

$$A = A_{22}A_{21}A_{12}A_{11} = (11101)(00101011)(11001010)(11010011011)$$

now, since their length is greater than the treshold, we need to split again all this substrings except $A_{22}$.
The final result of this operation on $A$ results in the following schema:

$$A = A_{22}A_{212}A_{211}A_{122}A_{121}A_{112}A_{1112}A_{1111} =$$

$$= (11101)(0010)(1011)(1100)(1010)(1101)(001)(1011)$$

and, after doing the same thing for $B$ we get this:

$$B = (101)(11001)(01010)(010)(0101)(0110)(1011)(1000).$$

The next phase is **Rearrange**: in this phase the first thing to do is to exchange the schemes that we computed during the previous phase, so we use on $A$ the divisions that we computed on $B$, and on $B$ the divisions that we computed on $A$. In order to easily perform this task, we can simply count the number of bits of each substring, for each input string of the function. For $A$ this list will be $A_{div} = [5, 4, 4, 4, 4, 4, 3, 4]$ (the first number of the list is 5 because the length of the first substring of $A$, $A_{22}$, is 5), and for the same reasons, $B_{div} = [3, 5, 5, 3, 4, 4, 4, 4]$. Once we have this lists it is easy to compute the new substrings-schemes: we start by fetching the first element from $B_{div}$, in this case 3, and so we count the first 3 bits (from left to right) in $A$: (111), this is now the new first substring of $A$. For the same reason, the second block of $A$ is (01001) and so on until we reach the end of the string: at the end we have

$$A = (111)(01001)(01011)(110)(0101)(0110)(1001)(1011)$$

and

$$B = (10111)(0010)(1010)(0100)(1010)(1101)(011)(1000).$$

The second task that the *Rearrange* phase performs, is the circular left shift of each substring through the *Rotation* function. Each substring $w$ of the input strings, is replaced by $Rot(w)$.

After we replaced every substring by the rotated one, we have the following configuration:

$$A = (111)(00101)(11010)(011)(0101)(1001)(0110)(1101)$$

and

$$B = (11011)(0100)(1010)(1000)(1010)(1110)(101)(0001).$$

We now use this two rearrangements of the strings to compute the final result of the function during the last phase: the **Composition**. This last phase simply consists in a XOR operation between the two outputs of the previous phase. From this operation, we get that with $T = 6$, the final result is:

$$Conv(A, B) = 001111111000011100001110001111100$$

## 1.3 Implementing The SLAP Protocol

In order to study the protocol, and find potential weaknesses, we may need to do some statistical analysis on the values that the Tag and the Reader exchange. Here, we propose a *JAVA* implementation of the protocol: by this implementation, every kind of test we could need in the future would be fairly easy to do.

### 1.3.1 Implementing The *Conversion* Function

The first thing to do, in order to implement the full protocol, is to implement the *Conversion* function. Here we have the code representing all the operations executed in order to compute the outcome of the function:

```
public String Conversion(int treshold, String A, String B){

    //GROUPING
    ArrayList<Integer> A_Division = Utility.computeDivision(A, treshold);
    ArrayList<Integer> B_Division = Utility.computeDivision(B, treshold);

    //REARRANGE
    ArrayList<String> rearrangeA = Utility.split(B_Division, A);
    ArrayList<String> rearrangeB = Utility.split(A_Division, B);

    rearrangeA = Utility.rotateAll(rearrangeA);
    rearrangeB = Utility.rotateAll(rearrangeB);

    String finalA = Utility.join(currentA);
    String finalB = Utility.join(currentB);

    //COMPOSITION
```

```
    return Utility.stringXOR(finalA, finalB);
}
```

The first thing to note, is the class `Utility`, which provides us with all the lower level methods that we need (you can find the code for the cited methods and others in the appendix). As we can see, for the *Grouping* phase, we need only the method `computeDivision(String input, int treshold)`. This is the most complicated method needed to implement *Conversion.* The aim of this method is to compute all the indexes in which the input string would be splitted, according to the rules of this phase. The method works as follows: first of all, an empty array of integers is defined; after that, it calls a recursive method that splits the input string into two substrings, and recursively splits all the substrings until the length is smaller or equal to the treshold. Once it finds a substring, whose length is not greater than the treshold, it takes the indexes relative to the location of the substring inside the first input strings, and checks if the indexes are inside the array defined at the beginning: if the indexes aren't in the array, then it adds them, otherwise, if they already are in the array, the method doesn't insert them again. It is necessary to check if an index has already been inserted, because, except for the indexes 0 and 32 (if the input string consists of 32 bits), the method would try to insert every index two times: once because it would be the last index of a substring, and another time because it would be the first of the following substring. The code for the `computeDivision` method is the following:

```
public  ArrayList<Integer> computeDivision(String input, int treshold){

    ArrayList<Integer> target = new ArrayList<>();
    Utility.computeDivisionAux(target, input, 0, input.length(),
                                                          treshold);
    return target;
}

private static void computeDivisionAux(ArrayList<Integer> target,
                       String input, int begin, int len, int treshold){

```

```
    if( (len - begin) <= treshold){
        if(!target.contains(begin))
            target.add(begin);

        if(!target.contains(len))
            target.add(len);
        return;
    }

    int w = Utility.getHammingWeight( input.substring(begin, len));
    Utility.computeDivisionAux(target,  input begin, len - w, treshold);
    Utility.computeDivisionAux(target, input, len- w, len, treshold);
}
```

Once this phase is done, we need to perform the two tasks of the *Rearrange* phase, so we need to split the strings and to left-rotate all the generated substrings. To split the strings, we implemented the method `split(ArrayList<Integer> indexes, String toSplit)`: it simply returns an array containing all the substrings generated by splitting, at the indexes indicated by the array, the input string `toSplit`. This method lets us split the strings only once, because we can directly split a string based on the indexes generated by the other one, so we don't need to split them two times, like one normally would do trying to calculate this function by hands. Once splitted, in order to perform the second task of the *Rearrange* phase, we need to replace each substring $w$ with $Rotation(w)$. This task is accomplished by the method `rotateAll(ArrayList<String> toRotate)`, that simply left-rotates all the strings in the array by their Hamming Weight.
At this point we are almost done: the method `join` trivially concatenates all the substrings together, according to the order in which the substrings were inside the first string, and, eventually, the `computeXOR` method returns the output of the third phase: the final result. As mentioned earlier, you can find the code fot these methods (and others) in the appendix.

### 1.3.2 Calculating the messages

As we said, the *Conversion* function is the most complicated task of all the implementation: once we implemented it, the messages will be fairly easy to compute. The first message to be computed by the Reader is $A = Conv(k_1^{new}, k_2^{new}) \oplus n$. From that formula we immediately obtain the function:

```
public String computeA(String k1, String k2){

    String n = Utility.getRandomString(k1.length());
    String c = Conversion(k1, k2);
    return Utility.StringXOR(c, n);
}
```

The only thing to note here is that the method `public static String getRandomString(int length)` returns a random binary string of arbitrary length.
The $B$ message is much more complicated than $A$, but, again, at this point we have all the necessary components and we only need to put them together. Recall that $B = Con(Rot(k_1^{new}, n), k_1^{new} \oplus k_2^{new}) \oplus Rot(Con(k_2^{new}, k_2^{new} \oplus n), k_1^{new})$. To have a more readable, organized, and maintainable code, we divide $B$ in the following way:

$$\mathtt{A} = Rot(k_1^{new}, n)$$

$$\mathtt{B} = Con(k_2^{new}, k_2^{new} \oplus n)$$

$$\mathtt{C} = Rot(Con(k_2^{new}, k_2^{new} \oplus n), k_1^{new}) = Rot(\mathtt{B}, k_1^{new})$$

$$\mathtt{D} = Con(Rot(k_1^{new}, n), k_1^{new} \oplus k_2^{new}) = Con(\mathtt{A}, k_1^{new} \oplus k_2^{new})$$

so that the result is trivially $\mathtt{C} \oplus \mathtt{D}$. We obtain the following code:

```
public String computeB(String K1, String K2, String n, int treshold){

    String A = Utility.rotate(K1, getHammingWeight(n));
    String B = Conversion(treshold, K2, Utility.stringXOR(K2, n));
    String C = Utility.rotate(B,  Utility.getHammingWeight(K1));
    String D = Conversion(treshold, A, Utility.stringXOR(K1, K2));
    return Utility.stringXOR(C, D);
}
```

The last message produced during the protocol is the message $C$, but, we'll not discuss an implementation of this message, since, as we'll see in the following chapters, this is completely useless to our purpose.

# Chapter 2

# Attacking The SLAP Protocol

Finding an attack to an authentication protocol of this kind, means to find the possibility to do, within the protocol, something that the protocol itself was built to deny. What if, for example, we find a way to update the values $k_1$ and $k_2$ on a Tag, from a device that is not the *authorized* Reader, preventing them from communicate again? Of course we don't want this to be possible, otherwise, why do we care about a secure autentication protocol? The study of this kind of protocols is necessary, in order to to be as safe as possible that it's impossible to do such things. So, for this kind of analysis, we need to be able to *think like an attacker*, without forgetting our good purposes.

## 2.1 The Attacker's Toolbox

The first necessary step, *thinking like an attacker*, is to be fully aware of the set of tools that an attacker could use to try to break a protocol. It's obvious that, in order to think about the ways an attacker would break a protocol, we need to embrace his power. The attacker-tools are basically three, and they are the following:

1. **Send:** This is trivial, but it's worth mentioning: an attacker can send messages to the various devices involved in the protocol, in this case, the Reader, and the Tag.

2. **Block:** This is not so trivial: an attacker can block a message sent from a device to another, preventing the second device to receiving it.

3. **Eavesdrop:** This is another powerful tool of which an attacker can take advantage: it can eavesdrop the messages that the devices send between them.

## 2.2 The Kinds of Attacks

There are various kinds of weaknesses, which can be found inside cryptographic protocols. There are also various kinds of way to exploit these weaknesses in order to attack a protocol. The ways to attack a protocol can be slightly different, and they're divided into several cathegories; here we present the foundamentals:

1. **Replay Attack:** This kind of attack consists basically in eavesdropping some values during a communication, and *replaying* them, trying to do other iterations of the protocol. Basically, referring to the $SLAP$ protocol, you can think of, for example, eavesdropping the messages $A$ and $B$ sent from the Reader to the Tag, and send them to the Tag again: in this case, nothing happens because the new values of $k_1^{new}$ and $k_2^{new}$ will be equal to the old values, and so they will be equal to the values stored on the authorized reader... but that is not the same for all the other protocols.

2. **De-synchronization Attack:** This kind of attack is one of the most common. It consists in *de-synchronizing* one device from the other, so that the two devices won't be able to communicate again. To be more clear, referring to the $SLAP$ protocol, a desynchronization attack would consist of a way to update the keys only to the Tag's side, so that the reader wouldn't be able to authenticate it anymore.

3. **Full-Disclosure Attack:** An attack is a *full-disclosure* attack when it lets us retrieve the secret values held by the various devices involved in the protocol. As before, thinking about the *SLAP* protocol, a full-disclosure attack would be an attack that, in some way, would let us recover $k_1$ and/or $k_2$.

4. **Impersonification Attack:** An impersonification attack is an attack in which the attacker pretends to be one of the devices involved in the protocol, and manages to carry out some operations that only the authentic device is supposed to execute.

## 2.3 An Attack to the *GUMAP* Protocols

In 2016 *Safkhani* and *Bagheri* [5] proposed a generalized *desynchronization attack* to the suit of the *GUMAP* protocols. This attack revealed some vulnerabilities in the common structure of all the *GUMAP* protocols, and, it was so important because, for a subset of protocols including *SLAP*, it was the first attack being published. Now we'll explain this attack with a pratical example of how it would work on *SLAP*.

Firstly, we assume that both the Reader and the Tag stores the two triplets of values $(k_1^{n-1}, k_2^{n-1}, ID_{n-1}) = (k_1^{old}, k_2^{old}, ID_{old})$ and $(k_1^n, k_2^n, ID_n) = (k_1^{new}, k_2^{new}, ID_{new})$. Let's see the steps for the attack:

**Session 1:** the Tag and the Reader complete a session of the protocol. In order to do that, they have to send the values $ID_n$, $A_n$ and $B_n$ through the insecure channel, so we *eavesdrop* and store them.

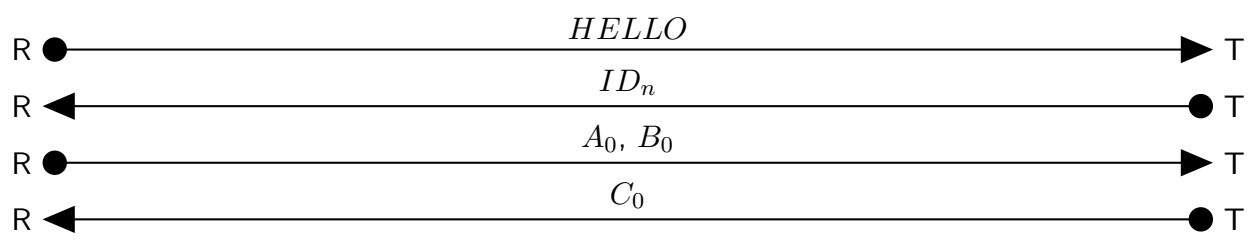

At the end of this session, the two triplets on both sides will be $(k_1^n, k_2^n, ID_n)$ and $(k_1^{n+1}, k_2^{n+1}, ID_{n+1})$.

**Session 2:** the Reader and the Tag complete the first steps of the protocol, until the messages $A_{n+1}$ and $B_{n+1}$ are sent from the Reader to the Tag: at this point, we *eavesdrop* all the values ($ID_{n+1}$, $A_{n+1}$ and $B_{n+1}$), and *block* $A_{n+1}$ and $B_{n+1}$, preventing the Tag from receiving it.

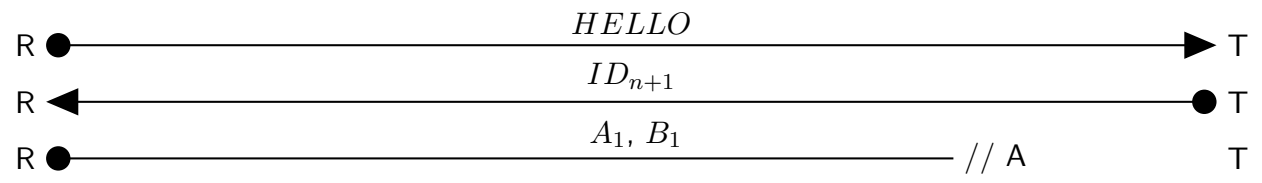

Since we blocked the communication, the protocol is not completed, and the values on Tag and Reader stays the same as before.

**Session 3:** the Reader sends an $HELLO$ message to the Tag, that sends back the value $ID_{n+1}$. At this point, we *block* the message sent by the Tag, and we *send* to the Reader (which is waiting for the $ID$) a random number. Since the Reader could not find a stored $ID$ corresponding to the random number, it will send another $HELLO$ to the Tag, which will respond with $ID_n$. At this point the protocol continues and both the devices will update their stored values.

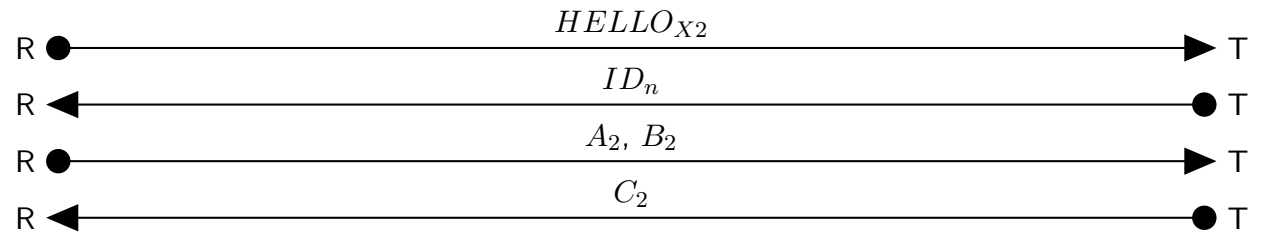

At the end of this session the values ($k_1^{n+1}$, $k_2^{n+1}$, $ID_{n+1}$) are replaced with ($k_1^{n+2}$, $k_2^{n+2}$, $ID_{n+2}$). It is important to notice, that, in this case, the triplet *new* has been replaced with the new computed values, but the triplet *old* hasn't been modified.

**Session 4:** during this session we *send* two $HELLO$ messages to the Tag, in order to let it send us the value $ID_n$. Once it sent the value, we send to the Tag the values $A_0$ and $B_0$ that we *eavesdropped* during session 1.

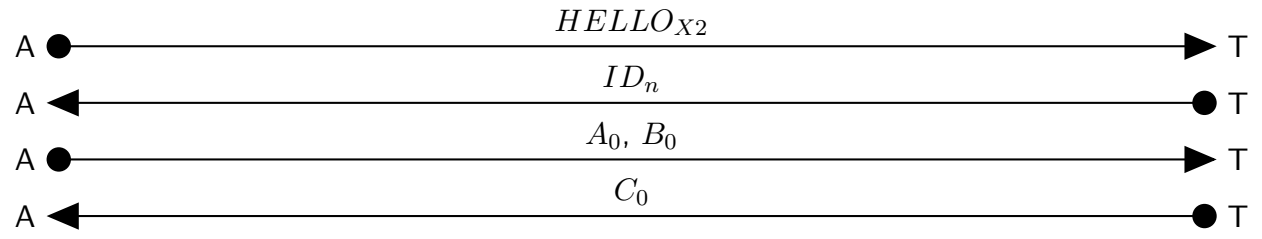

At this point, the Reader contains the two triplets ($k_1^n$, $k_2^n$, $ID_n$) and $(k_1^{n+2}, k_2^{n+2}, ID_{n+2})$; while the first triplet of the Tag has been modified. The Tag now stores $(k_1^n, k_2^n, ID_n)$ and $(k_1^{n+1}, k_2^{n+1}, ID_{n+1})$.

**Session 5:** we impersonate again the reader, and send the $HELLO$ to the Tag. This time, it will send $ID_{n+1}$ (because it is its $ID_{new}$). Now we send $A_1$ and $B_1$ that we *blocked* during session 2, we can because the authorized reader computed that values starting from $ID_{n+1}$.

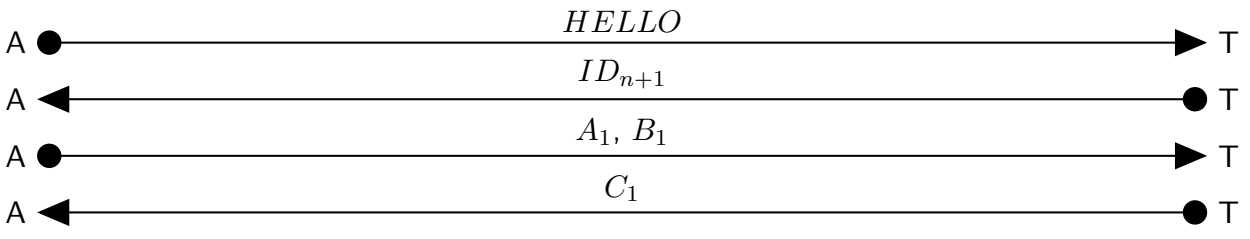

At the end of this session, the two triplets on the Reader are still $(k_1^n, k_2^n, ID_n)$ and $(k_1^{n+2}, k_2^{n+2}, ID_{n+2})$, but the triplets on the Tag are $(k_1^{n+1}, k_2^{n+1}, ID_{n+1})$ and $(k_1^{n+3}, k_2^{n+3}, ID_{n+3})$. It is easy to see that there isn't one common triplet stored on both devices, and so we desynchronized them.

## 2.4 Finding Vulnerabilities in SLAP

At this point, the functioning of the protocol has been explained in a great detail, furthermore, we also know some important weaknesses that let us understand the protocol even more deeply. With this understanding of the protocol, we are ready to look at it like an attacker would do. Since we have already seen an attack that only exploits a vulnerability inherent to the structure of the mechanism of $ID_s$ and keys updating, we will focus our attention on the other parts of the protocol, because our aim is to find new vulnerabilities.

### 2.4.1 Analysis of the Conversion function

As we said, the *Conversion* function is one of the main strengths of *SLAP*, so, finding a vulnerability in it would be a great starting point in order to carry out an attack on the protocol.

#### 2.4.1.1 *Inverting* the function

If we consider input strings of 32 bits, the number of possible outcomes is $\leq 2^{32}$, while, the number of the possible input couples, is roughly $2^{32} \times 2^{32}$. From this observation, we can easily conclude that, for nearly every outcome of the function, there is a group of input strings that produces it. If $A$, $B$, $C$ and $D$ are four strings of 32 bits, and $Con(A, B) = Con(C, D)$, we say that there is a collision between these two inputs and, since $Conversion$ allows collisions, we are sure that it is almost impossible to *invert* it. However, we can find a way to generate a couple of strings that produce a certain fixed output. The key observation is the following: suppose that $A$ is a 32 bits string, for example:

$$A = 01010111001000101111011010110010$$

and $B$ is a string of the following form:

$$B = 10000000000000000000000000000000$$

what happens if we try to compute $Conv(A, B)$ ?

Well, let's compute the outcome step-by-step: during the first phase (grouping), the string $B$ will generate the following splitting-schema:

$$B = |100000|0|0|0|0|0|0|0|0|0|0|0|0|0|0|0|0|0|0|0|0|0|0|0|0|0|0|$$

This schema will then be applied to the string $A$ during the second phase (rearrange), so, only the six leftmost bits of $A$ will be rotated.

Regarding the string $B$, it will be splitted according to the splitting schema generated by $A$, but, we only need to consider what happens in the leftmost block: in fact, except for the only 1 in the string, $B$ is only composed by $0s$, and so if we pick up a block that is composed by all zeros, they won't be rotated. In this case, the splitting schema generated by $A$ is the following:

$$A = |010|10111|00100|01|01|11101|10101|10010|$$

From this divisions, we can compute the output of the second phase (rearrange):

$$A = 10101011001000101111011010110010$$

$$B = 00100000000000000000000000000000$$

At this point, only the third phase (composition) is left, but we can easily conclude that the final result is:

$$Conv(A, B) = 10001011001000101111011010110010$$

We can observe that this result has an important characteristics: except for the six leftmost bits, it is identical to the string $A$. With some little changes, we can exploit this strange situation in order to produce, for every output $R$, two strings $A$ and $B$ such that $Conv(A, B) = R$.

We just saw that, if we act like in the previous example, no matter how we pick up the $A$ string, the outcome will differ from it (at most) for the six leftmost bits, both because these bits will be rotated during rhe rearrange, and because there will be the 1 in the $B$ string that will complement a bit in that area of $A$. First of all, suppose we know that, after the rearrange phase, the only 1 in $B$ would be the leftmost bit of the string: if we complement the leftmost bit of $A$, we would *neutralize* the modification of $A$ during the composition phase. As regards the rotation of the bits, we can easily avoid this problem because we already know the size of the block that would be rotated (and of course we know the Hamming Weight of the block). Since we have got these information, we can rotate the block to the right, by its Hamming Weight: in this way, the function will rotate back the block by the same number of positions to the left during the rearrange, recomposing it as it was before. After these two modifications, this is how $A$ looks like:

$$A = 01011111001000101111011010110010$$

Now, since we have supposed that the only 1 in $B$ is the leftmost bit during the composition phase, we can use the same method we used for $A$, to ensure that this happens: first of all we have to compute the size of the leftmost block of $A$ (as it is now):

$$A = 01|01111|1001000|101111011010110010$$

with this information, we know that the leftmost block of $B$ is $|10|$ and we can rotate it by one position to the right, obtaining our final $B$ string that looks like that:

$$B = 0100000000000000000000000000000$$

In order to check if we suceeded, let's compute the output of $Conv(A, B)$. This is the output of the grouping phase:

$$A = |01|01111|10010|00|10|11110|11010|110010|$$

$$B = 010000|0|0|0|0|0|0|0|0|0|0|0|0|0|0|0|0|0|0|0|0|0|0|0|0|0$$

this is the output of the rearrange phase:

$$A = 110101|11001000101111011010110010$$

$$B = 10|00000000000000000000000000000$$

and finally we have the output of composition:

$$Conv(A, B) = 01010111001000101111011010110010$$

As we expected, the result corresponds to the string $A$ that we picked at the beginning.
There could be many variations of this process, nevertheless, the procedure that we just saw, is the easiest one. The number of $1s$ of the string $B$ can be different from one: in fact, it only has to be $\leq 4$ (assuming that we have a *treshold* $\geq 4$). For example, by our setting, we could have picked up a string $B = 11100000...000$. When we choose a $B$ string with more than one 1, things gets much trickier, in fact, the output of the grouping phase of the $B$ string that we picked as an example would be:

$$|11100|000|000|000|000|000|000|000|000|000|$$

this means that the string $A$ woult be splitted into several blocks of three bits (except for the leftmost block) instead of blocks of only one bit, so, during the rearrange phase, these bits would be rotated. We can avoid this problem by rotating all the 3-bits blocks to the right, as we did before for the other blocks, the rest of the process stays the same, except for the fact that, this time, we (of course) have to complement the three leftmost bits in $A$.

The reason why we cannot use a string with more than four ones is not so straightforward. Suppose we want to pick up a $B$ string with five $1s$, here we have the related division-schema:

$$B = |11|11100|00000|00000|00000|00000|00000|$$

and suppose that, we have a certain string $A$ such that its division schema, transposed on $B$, looks like that:

$$B = |11|111000|000000000000000000000000|$$

according to our algorithm, we have to right-rotate the blocks, but, here's what happens: once that the bits have been rotated, the division-schema generated by the string changes!

$$B = |110|0011|10000|00000|00000|00000|00000|$$

so, if we use a string with five or more $1s$, the division-schema generated by $B$ after the right-rotation could change in an unpredictable way, and for this reason we cannot use such strings.

#### 2.4.1.2 Analysing collisions

Since we now know that it is impossible to invert $Conversion$, but we have a powerful algorithm to generate collisions (and we know that there are many couples of input strings that generate every output), we could wonder if these couples of input strings are related in some way.

In order to investigate along this path, we carried out some experiments. The first esperiment was the most natural for this purpose: we found some collisions and we looked for similarities among the couple of input strings. Here we have an example of collision:
input couple 1:

$$A_1 = 10011011001011000111000011011111$$

$$B_1 = 11110100101111100000011000011101$$

input couple 2:

$$A_2 = 10100110011110101111100101111101$$

$$B_2 = 00101000011111010101010111100001$$

final result:

$$R = 10010100001110011010000001111000$$

Like in this case, we had to look for some common features between the couples: since at first glance we couldn't identify noticeable similarities, we tried to consider some specific parameters. As parameters, we chose to evaluate the same parameters considered by the function itself: Hamming Weight and locations of different bits. Let's take a look to the Hamming Weights of these strings: $wt(A_1) = 18$, $wt(B_1) = 21$, $wt(A_2) = 17$, $wt(B_2) = 16$. From this comparison, we cannot identify noticeable similarities. Another interesting comparison consisted of locating the differences between members of the first couple and members of the second couple, here is an example:

$$A_1 = 10011011001011000111000011011111$$

$$B_2 = 00101000011111010101010111100001$$

$$XOR_1 = 10110011010100010010010100111110$$

We have tried to analyse the differences between all possible pairing of strings for every collision and we have tried this experiment on 100 collisions. Unfortunately, this experiment hasn't revealed something of interesting.
Another idea, has been to fix two strings $A$ and $B$, and try to apply little changes to them, in order to check if we could obtain some strings $A'$ and $B'$ such that $Conv(A, B) = Conv(A', B')$. The goal was the same of *section 2.4.1.1*, but, this time, we used $A$ and $B$ as our starting point, instead of using the result. Starting from this idea we designed two similar experiments: in the first experiment, we changed a single bit in $A$, and we computed $Conversion$ by flipping each time a different bit in $B$. We repeated this for every bit in $A$, in order to check all possible combinations of one flipped bit in $A$ and one flipped bit in $B$. The second experiment prescribed to do the same thing, but each time flipping two bits in $A$ and two bits in $B$, in all possible ways. From a practical point of view, this was an inefficient procedure, but it was feasible for strings of 32 bits (it was almost impracticable for strings of 96 bits, since it would require $\binom{96}{2} \times \binom{96}{2}$ trials), but, we were interested in finding vulnerabilities, and this was just a procedure that could have given us a starting point.
After the execution of both tests, the results said us that, in both cases, the probability to get a collision in this way is $\approx 0.5\%$. It is a very discouraging result and it definitely told us that this was not the right place to look for vulnerabilities, even though, we can move forward with a deeper understanding of $Conversion$.

#### 2.4.1.3 How one-bit flip affects the output of *Conversion*

In the previous section we realized that, starting from some input strings, it is almost impossible to generate a collision. This phenomenon is not accidental: it relies on the cryptographic properties of *diffusion* and *confusion*. These two properties indicate the fact that, for a certain cryptographic function $f$, no matter how two inputs $x_1$ and $x_2$ are similar, $f(x_1)$ and $f(x_2)$ have to show no relations at all between them: they have to seem randomly picked from the set of the possible outputs. The results of the previous experiments suggested that the creators of *SLAP* have succesfully built a function with these characteristics. Anyway, the tests that we carried out before were too specific. Starting directly from the definition of *diffusion* and *confusion*, we designed a test that is specific for this purpose. The test consisted in the following steps: we chose two random strings $A$ and $B$.

At first, we computed the output string $r = Conv(A, B)$. After that, we randomly selected a bit in $A$, we flipped it and we called the so obtained string $A'$. Again, we computed $r' = Conv(A', B)$, and we analysed the differences between $r$ and $r'$ in terms of how many bits changed, and where this different bits were located. As a second step, we obtained the string $A''$ by flipping one randomly selected bit inside $A'$, making sure that $A'' \neq A$. At this point, we computed $r'' = Conv(A'', B)$ and we compared the two strings $r'$ and $r''$. We repeated this test for 100'000 iterations on input of 32 bits and treshold = 6.

The diagram below shows the frequency of the number of bits that changed between one output and its next: for example, in the 6.40% of the cases, the number of different bits between one output and its next was 5.

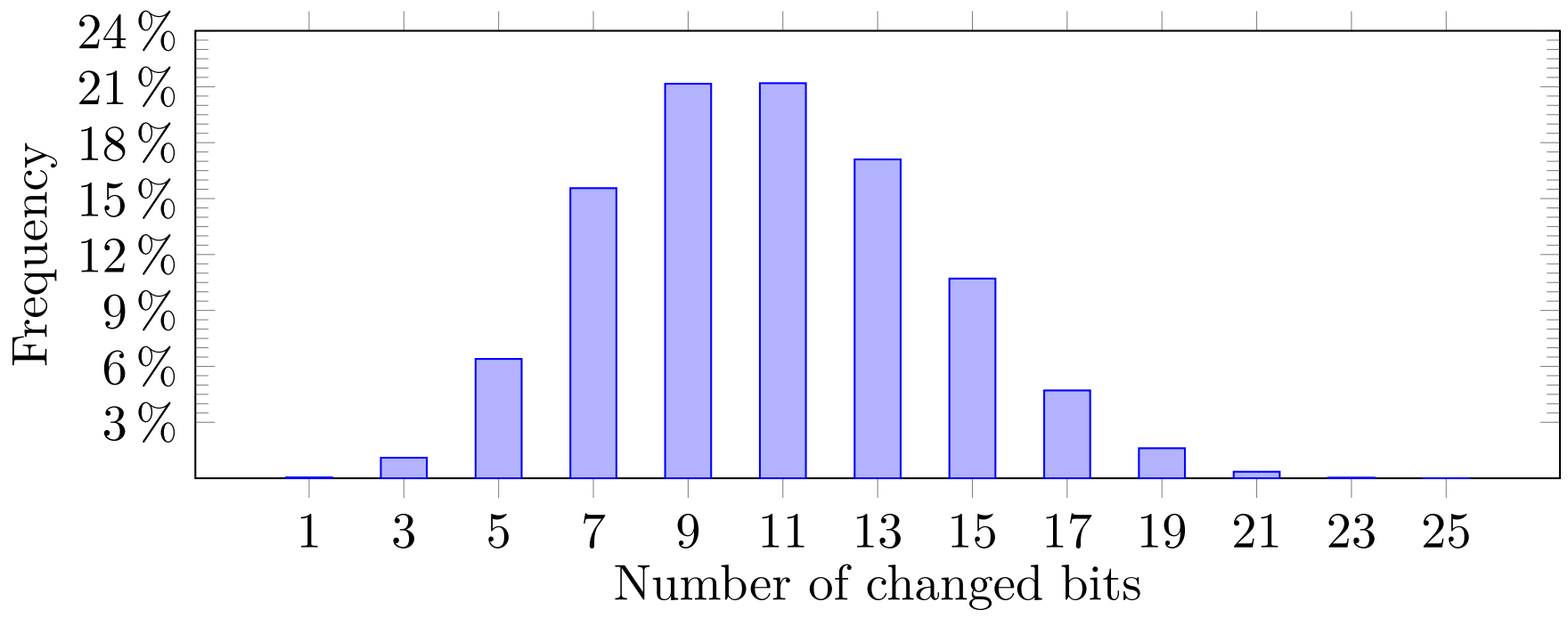

Already at a first glance, there is an important information that we can notice: the number of differences is always odd. This seems to be a structural characteristic of the function, and it seems the most important information that we can get from these results. Furthermore, as we can see from the diagram, the number of differences between one output and the next follows approximately a bell curve. This indicates that, under this aspect, the *Conversion* function shows a good behavior, in fact, the number of differences that we get each time follows an apparently random trend: even under this aspect, *Conversion* seems to show that the properties of *confusion* and *diffusion* have been well implemented.

Regarding the positions of this different bits, their behavior seems to be completely random, and it doesn't let us make any assumption.

#### 2.4.1.4 How 2-bits filp affects the output of *Conversion*

This test was very similar to the previous, but, this time, instead of flipping one single bit per iteration, we flipped two. Even this time, the positions of the bits to flip was randomly selected in $A$, and, of course, we made sure that each time $A'' \neq A$.

As we did for the previous experiment, we summarized the results in a diagram:

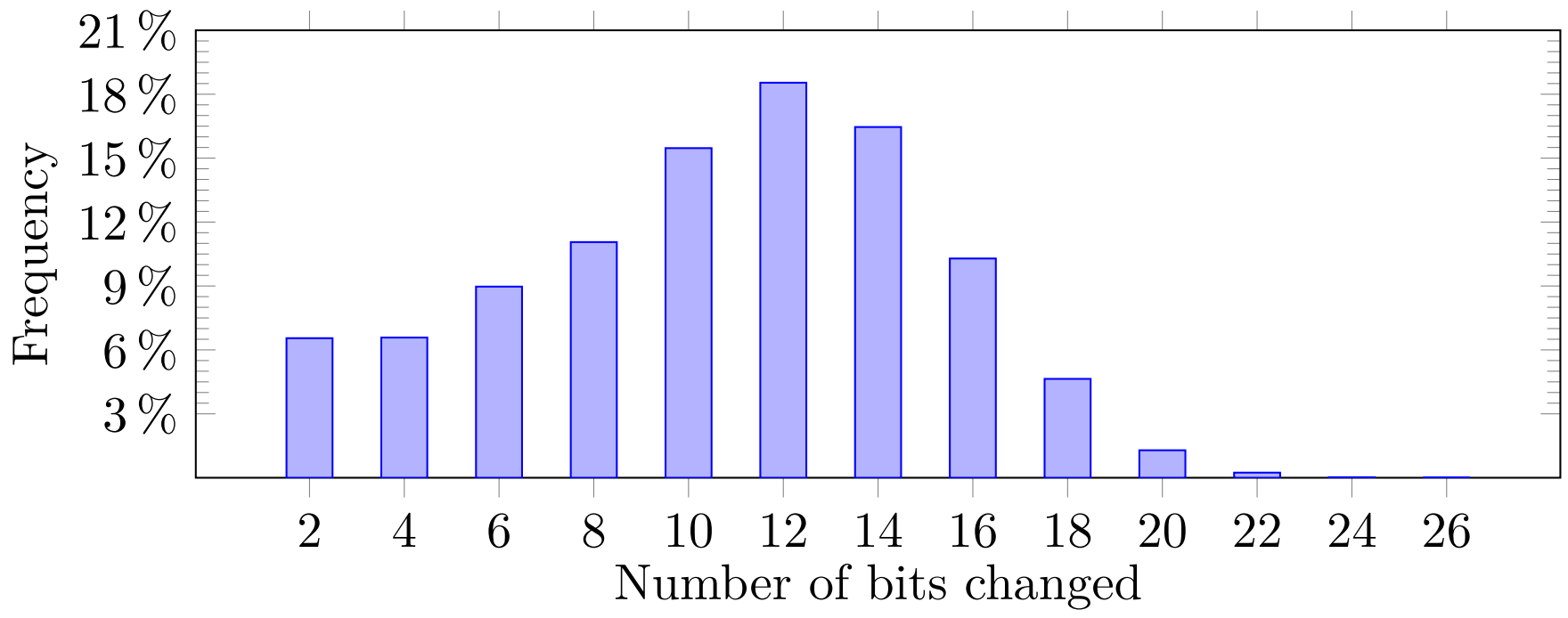

Even this time, we can immediately notice something interesting: the number of differences is always even. In the case of the previous experiment, we were modifying the Hamming-Weight of the string by an odd factor, and the differences between the outputs were odd; in this case, the modification of the H-W has been by an even number of bits, and it resulted in an even number of different bits between the outputs. As soon as we noticed this pattern, we repeated the same experiment for modifications of 3 and 4 bits, in order to check if the outcomes would still present it. As we expected, the outcomes presented the same pattern, so, it appears reasonable to us to assume that it is generally true for every nomber of modified bits.

Another observation that we can make is that, in this case, even though we can still identify a sort of bell curve, it is slightly shifted to the left. We conducted a deeper analysis, taking into account even the positions of the flipped bits, and the positions of the different bits in the output. We could notice that there was a strange (and probably unexpected) behavior of the function: when the flipped bits were both located at an extreme of the string, say, the first six or the last six positions, and the two flipped bits were different from each other, the number of differences between the two results dramaticly decreased. We repeated the same experiment with different configurations of string length and values of the treshold parameter, flipping couples of two different bits both within the first of last six positions of $A$ in every possible way. The four tables of the following page summarizes the result.

From a theoretical point of view, this happened because if the bits were flipped at an extreme side, and the weight of the string was the same as before (and this is the reason why we wanted the two bits to be different), with great probability the division's schema generated during the rearrange phase wouldn't change, so the changes would have an impact only on the specific block in which the modified bits were located.

*How 2-bits filp affects the output of Conversion:*

Table 2.1: *treshold* = 6

| #Different bits | Frequency |
|---|---|
| 2 | 37% |
| 4 | 10% |
| 6 | 7% |
| 8 | 7.5% |

Table 2.2: *treshold* = 7

| #Different bits | Frequency |
|---|---|
| 2 | 40% |
| 4 | 6.9% |
| 6 | 5.8% |
| 8 | 6.42% |

Table 2.3: *treshold* = 8

| #Different bits | Frequency |
|---|---|
| 2 | 40.2% |
| 4 | 6.85% |
| 6 | 5.95% |
| 8 | 6.56% |

Table 2.4: *treshold* = 10

| #Different bits | Frequency |
|---|---|
| 2 | 45.4% |
| 4 | 1.45% |
| 6 | 3.6% |
| 8 | 6.1% |

### 2.4.2 Tests on the values $A$ and $B$

We noticed from the tests that we carried out on the *Conversion* function that, if we modify the inputs in some particular points, the consequences of the modifications on the output are minimized. Starting from this observation, we conducted some tests on the values that the Tag and the Reader exchange, in order to see how this strange phenomenon affected them. First of all, let's recall how the values $A$ and $B$ are computed:

$$A = Conv(k_1^{new}, k_2^{new}) \oplus n$$

$$B = Conv(Rot(k_1^{new}, n), k_1^{new} \oplus k_2^{new}) \oplus Rot(Conv(k_2^{new}, k_2^{new} \oplus n), k_1^{new})$$

As we can notice from the description of the protocol, the value $A$ is only used to hide the random number $n$, that will be used by the Tag in order to compute the new $IDs$. A more subtle observation is that, if we send a modified $A$ to the Tag, our changes would only affect $n$, because the Tag already knows $Conv(k_1^{new}, k_2^{new})$, and it will retrieve $n$ by simply XORing this value with $A$. Let's now suppose that we send a modified $A$ to the Tag: how will the changes we made in $n$ affect the $B$ value computed by the Tag?

Even though $B$ is calculated in a deliberately complex way, this doesn't prevent us from doing a simple theoretical analysis. There are two points in which $n$ is used in order to compute $B$: the first is $Rot(k_1^{new}, n)$, and the second is $Conv(k_2^{new}, k_2^{new} \oplus n)$. In our previous tests on *Conversion*, we observed that in order to get an output that has the smallest number of differences compared to the output obtained without modifications, a good idea is to modify the input in two bits, that are one different from the other, in this way Hamming Weight will stay the same. The situation is similar for $B$, because one of the two expressions in which $n$ in involved is $Rot(k_1^{new}, n)$: in this situation, if the weight of $n$ stays the same, the result of the expressions won't be modified at all.

As regards the second situation in which $n$ is used, $Conv(k_2^{new}, k_2^{new} \oplus n)$, it is pretty much the same: if we modify two input bits located at an extreme side of the input of *Conv*, we can expect its output to be pretty similar to the output we would get without the changes, except for a very small amount of bits located at such side. In order to understand if our analysis is correct, we conducted some practical tests on the values $A$ and $B$ as we did for $Conversion$.

#### 2.4.2.1 How 2 bits-flip in $n$ affects $B$

For our first tests, although we took into account our previous considerations, we tried to be as more general as possible. The first test has been carried out in this way: we generated a random suit of all the necessary values: $k_1$, $k_2$ and $n$, and then, starting from them, we computed a first $B$ value. After we computed the first $B$, we flipped a couple of bits in $n$, we generated a new value $B'$ and we evaluated the differences between $B$ and $B'$ as we did for $Conversion$. We repeated this test for all possible couples of bit in $n$. There is an important difference between this test and the ones regarding $Conversion$: in this case, every value $B'$ has been compared to the first $B$: a comparison between two values $B'_1$ and $B'_2$ would have made no sense.
At his point, it is important to recall that not all $B$ is sent to the Tag: once that the value has been computed, its Hamming Weight is evaluated and then, if it is even, the right half of $B$ is sent to the Tag, while, if it is odd, the Reader sends to the Tag the left one, so, at the end, we will make considerations only on half of the string. We generated ten different suits of values $k_1$, $k_2$ and $n$, and performed the test on all of them.

The following diagram shows a mean of the results of the ten tests:

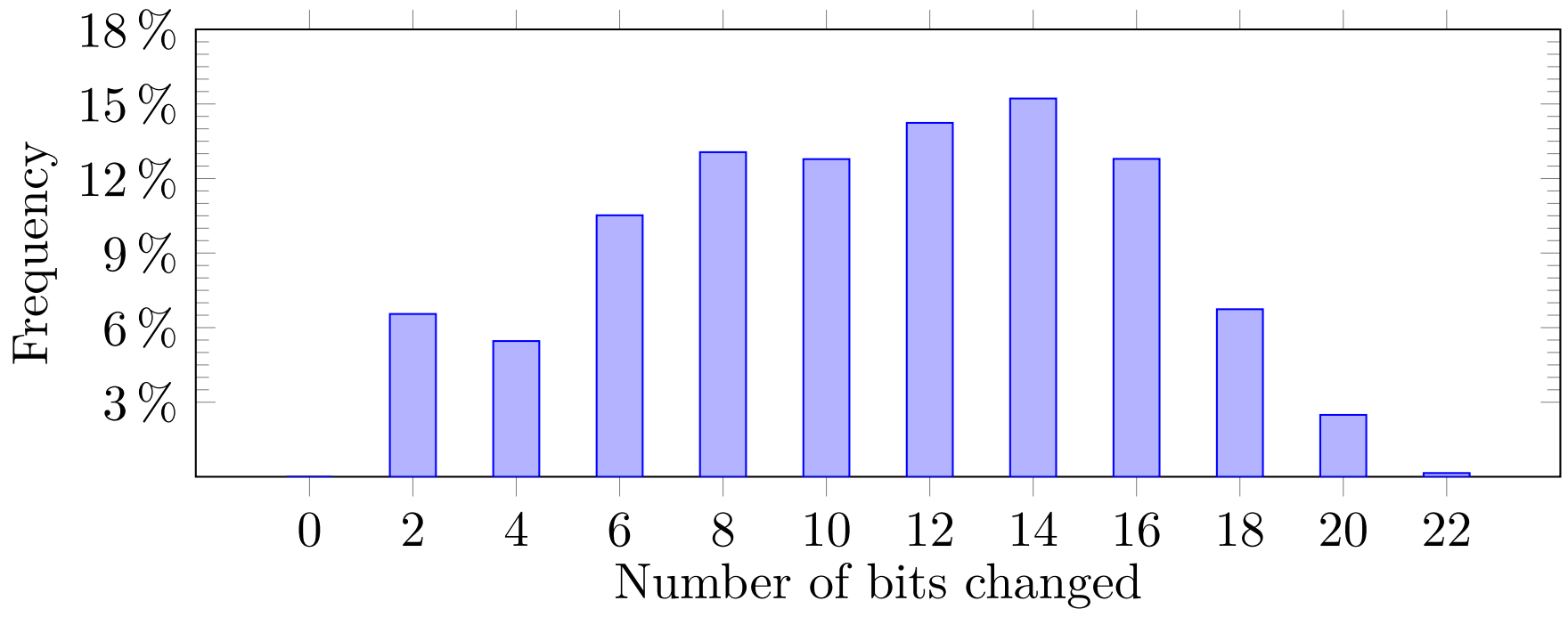

Let's analyse what these numbers means. First of all, we can notice an in-teresting and probably expected behaviour: as it happens for *conversion*, when we modify two bits in the input (in this case $n$), the output is modified by an even number of bits. A second observation that could be quite interesting is that, in none of the ten tests, even flipping two bits in all possible ways, we didn't found any collision: neither one!

In this case, to analyse only these aspects, such as the number of different bits, and their positions, would be reductive. As we said, the $B$ value that is evaluated on the Tag is only half of the entire value, and which half will be sent depends on the Hamming Weight of the entire string $B$, so, the Hamming Weight is an important factor and we need to consider it. Looking at how the Hamming Weight of these outputs changes, we noticed something interesting: it always varies for an even factor! This is so much interesting because, if we have a value $A$ and its associated $B_{L/R}$ we are sure that, if we modify $A$ in exactly two points, and we send it to the Tag, the Tag will compute $B$ and choose the same half (left or right) that the Reader chose.

#### 2.4.2.2 How changes in $n$ affects $B_{R/L}$

At this point, we already have some important information on the way certain changes in $n$ impact the final result $B$: if we modify $n$ in exactly two points, the new $B$ value will differ from an even number of bits, and also the differences between the two Hamming Weights will be even, so the half of $B$ that will be sent/evaluated won't change. These information tells us that the brief theoretical analysis that we presented before was right. The last implication of that general analysis was that, if we modify two bits at an extreme of $n$, the number of differences will be minimized.
This time we repeated the test, but, instead of fixing the $B$ value, we fixed only the corresponding half. As before, we flipped every possible couple of bits inside $n$, and each time we compared the output of the iteration with the "original" half of $B$ that we fixed at the beginning. We repeated the same test for ten suits of $k_1$, $k_2$ and $n$, and then we computed a mean of the results. Here we present, in the usual way, the results:

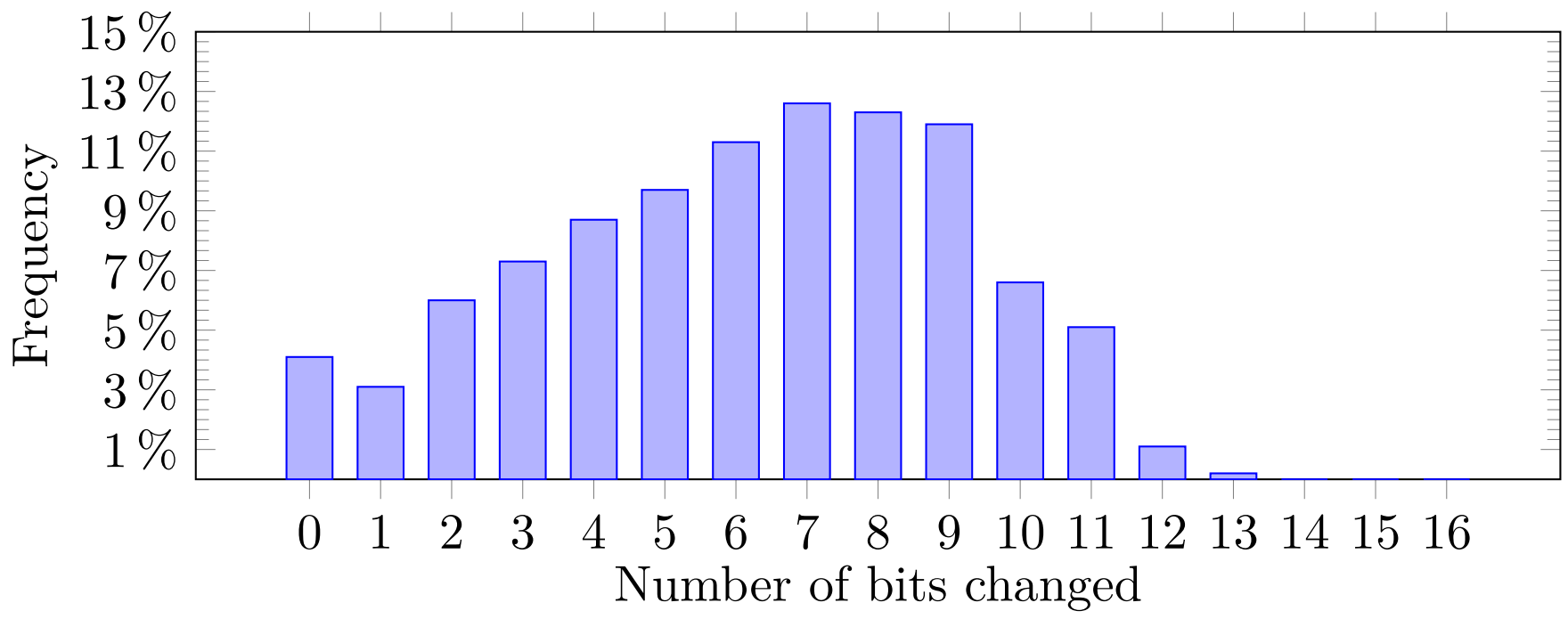

Probably, the most interesting result from these data, is the fact that there is a certain number of cases, about the 4.1%, in which there is a collision, that is, $B'_{R/L}$ (the one generated with the modified $n$) and $B_{R/L}$ are equals. This sort of behavior is certainly unintended: it tells us that, in certain cases, if we eavestrop a session, and we send a couple of values $A'$ and $B$, there is a probability that the Tag will autenticate us. It is true that the 4.1% of probability is too low for a practical purpose, however, this is a good starting point.

This diagram looks very similar to the diagram in *section 2.4.1.4*, it is "shifted to the left". Looking at the sequence of values $B'_{L/R}$ generated by the software that performed our test, we immediately noticed that there was a good number of collisions in correspondence of the very first or very last indexes. Considering that the cases of collisions were just a few, compared to the length of the sequence of $B$ values, the fact that there was a good amount of collisions located at some specific indexes, was a strong indication. And this was what we were expecting based on our previous observations.

### 2.4.3 Looking for collisions in $B_{R/L}$

Knowing that there are some collisions, our aim was to locate them with the utmost possible precision. As usual, we selected a suit of $k_1$, $k_2$ and $n$, and we computed $B_{R/L}$. Secondly, we selected a couple of bits from the first six or the last six positions of the string $n$, we flipped them, and then we computed a new value $B'_{R/L}$. We did the same thing for all possible couples of bits in the first six or in the last six positions of $n$, and each time we compared the output value with the first one.
We repeated this experiment for 100.000 sets of $k_1$, $k_2$ and $n$, for many configurations of string length and value of the treshold parameter. The table on the next page shows the results.

*Looking for collisions in* $B_{R/L}$*:*

| string length | treshold | collision found |
| --- | --- | --- |
| 32 | 6 | 83.42% |
| 32 | 7 | 83.2% |
| 32 | 8 | 83.5% |
| 32 | 10 | 84.62% |
| 96 | 6 | 64.04% |
| 96 | 7 | 64.05% |
| 96 | 8 | 63.81% |
| 96 | 10 | 61.82% |

## 2.5 An impersonification attack to the SLAP protocol

Thanks to the results we obtained in the last sections, we are now able to carry out an impersonification attack to $SLAP$ (remember that we defined this knid of attack in *section 2.2*). Through this attack, we will be able to carry out an unhautorized session of the protocol, and hence the Tag will update one of its triplets of values.
The *figure 2.1* summarizes the steps needed in order to carry out the attack: the first step is to *eavesdrop* a regular session between Tag and Reader; in particular, we are interested in the values $A$ and $B_{LorR}$. Starting from the eavesdropped $A$, we compute a value $A'$ by selecting two different bits, both located within the first six positions or whitin the last six positions of the string, and fliping their values, as we did for our last tests. As we know, for certain values of $A'$, the corresponding value $B_{LorR}$ computed by the Tag will be the same of the previous session. Therefore, if we choose the right $A'$, and we send the pair $A' - B_{LorR}$ to the Tag, it will authenticate us and update its triplet ($ID_{new}$, $k_1^{new}$, $k_2^{new}$) with a new set of values that is completely unknown to the authorized Reader. We need a few attempts in order to guess the right $A'$, but, once we guessed it, the Tag will notify us by sending $C_{LorR}$, which is the proof that the session has been succesfully completed. If we try all the possible $A'$, that means $\binom{6}{2} \times 2 = 30$ total attempts, the overall probability of success is the same of the experiment of *section 2.4.3*, so, it is between 63-73% (for each authorized session that we eavesdrop) depending on the combination of string length and treshold.

*An impersonification attack to the SLAP protocol:*

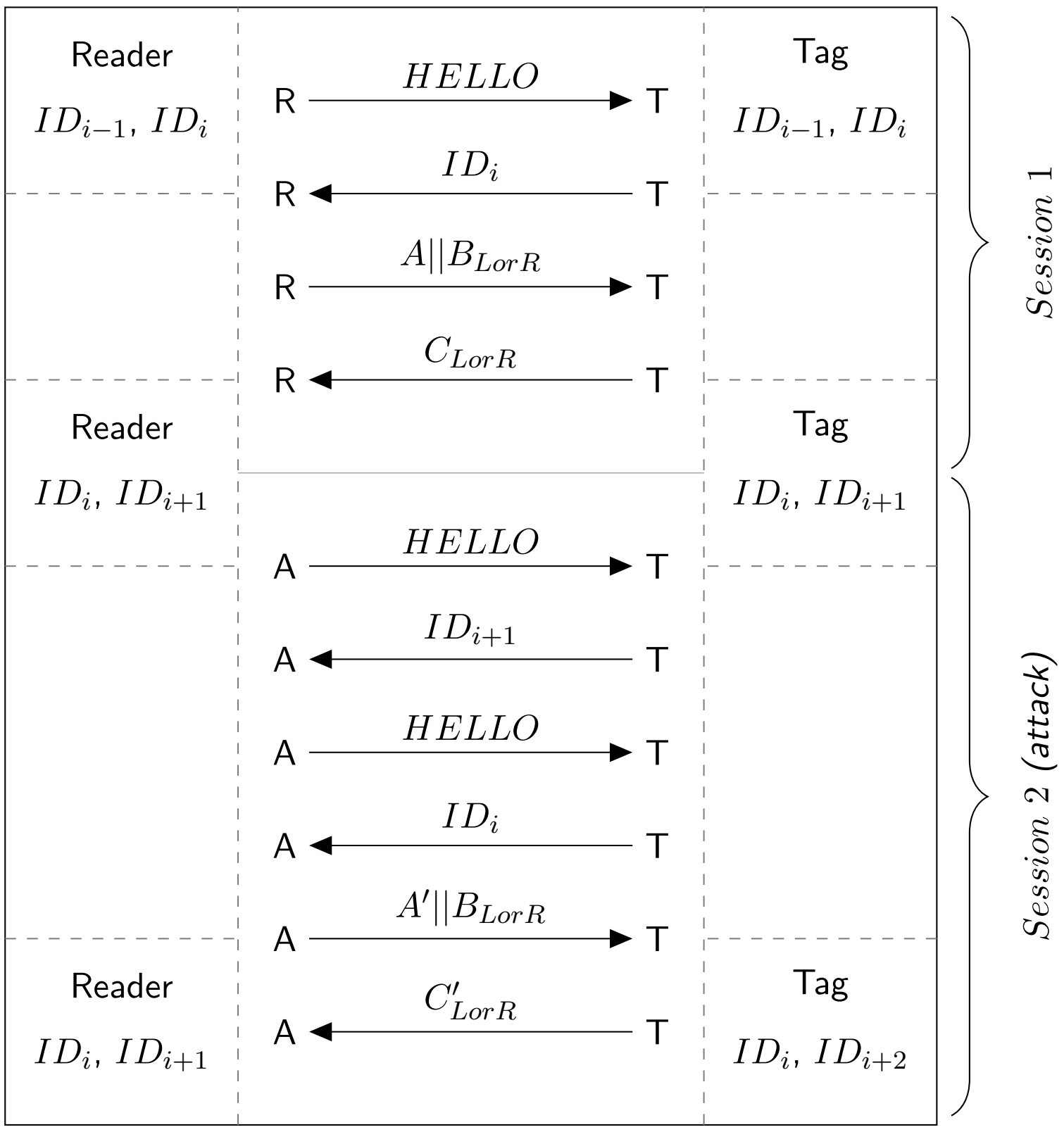

Figure 2.1: An impersonification attack to SLAP

### 2.5.1 Details for an efficient implementation

#### 2.5.1.1 Reducing the number of attempts

Due to the Ultra-Lightweight nature of this protocol (and therefore to the poor computational power of the Tags), we should try to use as less attempts as possible, in order to choose the right $A'$. First of all, flipping bits that are exactly located in the first (or the last) six positions is useful because, if we reduce this number of positions to five, our success rate falls down by about 10%, in case of 32-bits inputs, and by about 5.5% in case of 96 bits input (but we reduce the number of attempts to $\binom{5}{2} \times 2 = 20$). On the other hand, if we increase the positions to flip by another, the success rate only grows by about 2.5%, in case of 32-bit inputs, and by 1.5% in case of 96 bits input (but the number of attempts becomes $\binom{7}{2} \times 2 = 42$). For a protocol of this kind, our 30 attempts are still too much. Here we try to reduce them to a feasible number, but trying to maintain the highest possible success rate. Firstly we can assume that the half of $B$ that is sent to the Tag is always $B_R$ (i.e. the weight of $B$ is always even). This happens only in $\frac{1}{2}$ of the cases, but we reduce the number of attempts to 15. Since we reduced the number of attempts, the success-rate will be reduced too, *Table 2.5* shows the success rate if we only flip bits in the left side of the string. As a second step, we can try to identify if there are some flips that have an higher success rate compared to others. In order to understand if this intuition was right, we designed a simple test: we counted for each of our 15 flips, how many matches it would generate on 500,000 sets of $k_1$, $k_2$ and $n$. As we can see from *Table 2.6* of the following section, the success rate of a flip is not random, as one could expect, but, it follows a roughly defined classification, that is not affected very much from the length of the strings or from the value of *treshold.* Now that we know that there are some flips that have a relatively low success-rate, we can try to avoid them, further reducing the number of attempts required. If we exclude the last 5 attempts (0-5, 2-4, 1-5, 3-5, 2-5) the probability of *Table 2.5* reduces by 3% in each case: this is not so bad, because we reduced the attempts required to 10, and we still have a success rate of about 44%.

*Reducing the number of attempts:*

| String length | Treshold | Success rate |
|---|---|---|
| 32 | 6 | 47% |
| 32 | 7 | 47.7% |
| 32 | 8 | 48.12% |
| 32 | 10 | 48.22% |
| 96 | 6 | 47.33% |
| 96 | 7 | 47.27% |
| 96 | 8 | 48.22% |
| 96 | 10 | 47.5% |

Table 2.5: Success rate flipping only in the first six positions from left.

#### 2.5.1.2 Choosing an efficient strategy

Since now we know that some flips have an higher probability to produce a match, we can think of a strategy through which the various flips should be tried. Here we present two trivial strategies, and a third one, that is based on the classification in *Table 2.6* that you can find on the next page. As we can see, the five flips with the lowest probability have been excluded from the strategy, for the reasons we presented in the previous section.

**Strategy 1**:

(0, 1)-(0, 2)-(0, 3)-(0, 4)-(1, 2)-(1, 3)-(1, 4)-(2, 3)-(3, 4)-(4, 5) average attempts required = 3.95

**Strategy 2**:

(0, 1)-(1, 2)-(2, 3)-(3, 4)-(4, 5)-(0, 2)-(1, 3)-(0,3)-(1, 4)-(0,4) average attempts required = 3.55

**Strategy 3**:

(0, 1)-(1, 2)-(0, 2)-(2, 3)-(4, 5)-(3, 4)-(1, 3)-(0, 3)-(0, 4)-(1, 4) average attempts required = 3.45

As we expected, our *best* strategy performs better than the two trivial ones, though, it doesn't improve significantly the overall efficience of the attack. From these data, we know that, with ≈ 44% probability, on average we need only 3-4 attempts in order to succeed.

*Choosing an efficient strategy:*

| Flipped indexes | success rate |
|---|---|
| 0 1 | 11.7 % |
| 1 2 | 10.8 % |
| 0 2 | 10.7 % |
| 2 3 | 10.3 % |
| 4 5 | 10.2 % |
| 3 4 | 10.2 % |
| 1 3 | 10.0 % |
| 0 3 | 9.9 % |
| 0 4 | 9.5 % |
| 1 4 | 9.4 % |
| 0 5 | 9.3 % |
| 2 4 | 9.1 % |
| 1 5 | 9.0 % |
| 3 5 | 8.8 % |
| 2 5 | 8.7 % |

Table 2.6: Absolute success rate of the single flips.

### 2.5.2 Considerations about the attack

The result that we obtain with $\approx 44\%$ probability and few attempts, could be easily achieved with one attempt and almost 100% probability by performing the two following steps:

- eavesdropping the values $A$ - $B_{LorR}$ sent by the Reader to the Tag and blocking them.

- sending the eavesdropped values to the Tag.

By performing these two steps, we would let the Reader and The tag in the same condition we would let them by performing our attack. Even though this is absolutely true from a theoretical point of view, these two procedures produce very different situations in practice. The strength of the proposed impersonification attack, lays in the fact that it could easily and effectively be used in practice: all we need to do is to eavesdrop a session. As regards the procedure above, the situation is different because it requires to block a session, and, in the real world, the person/device that is trying to carry out a certain session will notice if the session has been blocked. Once noticed that the session has been blocked, one could take some countermeasures, for example, we could perform two consecutive sessions of the protocol, in this way the set of values that the attacker could have used in order to carry out the attack would disappear.
Even though there are many attack that from a theoretical point of view are much more powerful than the one proposed in this work (for example the desynchronization attack proposed by *Safkhani* and *Bagheri* [5]), they require to eavesdrop and block more consecutive sessions, and this could be not so easy in the real world, as it could seem on a piece of paper.
As regards the cryptographic property that this attack violates, it is the *Integrity.* Quoting the authors of the protocol: "...it is essential to ensure the received random number in the tag side is the same as the random number generated by the reader. *For the sake of this goal, the messages* $B_{LorR}$ *and* $C_{LorR}$ *not only provide the evidence for mutual authentication, but also assure the integrity of the transmitted messages. For example, when* *a malicious adversary tries to modify* $n$ *by flipping certain bits in* $A$*, it must transfer the result of random number* $n$ *which extracts from* $A$ *by the tag. Finally, the local computed message* $B'$ *at tag side will be incorrect and lead a failed authentication.* [1]"

#### 2.5.2.1 Possible fixes

In order to fix this vulnerability, we would need to spread the small changes in $n$ all along the string $B$. An expression that could help us to reach this goal could be

$$F = Con(Rot(n, A), A)$$

This expression has some interesting characteristics: first of all, we know that the value $A$, will differ from two bits (located at an extreme side) between the two sessions of the protocol involved in the attack. The same thing happens with $n$, but, since we expect $weight(A) \approx \frac{length(n)}{2}$, the modified bits would be shifted to an area of the string that certainly won't be an extreme side, by the expression $Rot(n, A)$. It is important to notice that, since the weight of $n$ must stay the same for the attack to succeed, and we cannot directly flip bits in $A$ with this aim, in about half cases, when the attack succeeds, the weight of $A$ will be modified (and the *Attacker* won't be able to succeed, if he pretends that the weight of $A$ stays the same, because his probability to succeed would decrease significantly). This is an important property that could help us to make the protocol safer.

Even small changes in $n$ would affect in a significant way the expression $F$, and, if we replace the value $B$ involved in the protocol, by the value

$$B^F = B \oplus F$$

the frequency of success of the attack would decrease significantly, making the attack useless for any practical purpose. It is important to notice that the introduction of $B^F$ would not require a significant amount of additional computational power to the Tag, that could compromise the Ultra-Lightweight nature of SLAP. The following tables show how the success probability of our attack would change if we implement the proposed fix.

Even though $\approx 4\%$ could seem a too high probability, we have to notice that it would require all the 30 attempts proposed at the beginning, so, it is clear that the attack would be useless once this fix has been implemented.

*Possible fixes:*

Table 2.7: Success rate considering all 30 flips

| Length | Treshold | Success |
|---|---|---|
| 32 | 6 | 4.0% |
| 32 | 7 | 4.3% |
| 32 | 8 | 4.9% |
| 32 | 10 | 4.2% |
| 96 | 6 | 4.0% |
| 96 | 7 | 2.7% |
| 96 | 8 | 3.9% |
| 96 | 10 | 2.7% |

Table 2.8: Success rate considering only the 15 flips in the right side

| Length | Treshold | Success |
|---|---|---|
| 32 | 6 | 2.2% |
| 32 | 7 | 2.3% |
| 32 | 8 | 2.6% |
| 32 | 10 | 2.3% |
| 96 | 6 | 2.9% |
| 96 | 7 | 2.7% |
| 96 | 8 | 3.9% |
| 96 | 10 | 2.7% |

# Appendix A

# The Utility class

In this appendix you will find some useful methods that we put into the `Utility` class. The aim of this class is to provide lower level functionalities, in order to enhance the code reusage, and let us concentrate on higher-level tasks.
The first method is used in order to get the Hamming-Weight of a string:

```
public static int getHammingWeight(String s){
    if(s == null)
        return -1;
    int count = 0;

    for(int i = 0; i < s.length(); i++){
        if(s.charAt(i) == '1')
            count++;
    }
    return count;
}
```

This second method is used in order to compute the $XOR$ of two strings:

```
public static String stringXOR(String A, String B){
    if(A == null || B == null || A.length() != B.length())
        return null;

    String result = "";

    for(int i = 0; i< A.length(); i++){
```

```
        if(A.charAt(i) == B.charAt(i))
            result += "0";
        else
            result += "1";
    }
    return result;
}
```

The `rotate` method is used to left rotate a string $X$, by $w$ positions:

```
public static String rotate(String X, int w){
    if(X == null || w < 0 || w > X.length())
        return "";

    if(X == null || X.length() == 0 || w < 0)
        return "";

    if(w == 0)
        return X;

    String p1 = X.substring(0, w);
    String p2 = X.substring(w);

    return p2 + p1;
}
```

The method `flipBit` is used in order to flip a specific bit of a string. Later in this appendix, you will find the method `flipAdjacentBits(int f, int e, String s)`. This method simply flips all the bits of `s` from the one with index `f` (included), to the one with index `e` (excluded).

```
public static String flipBit(int index, String s){

    if(index < 0 || s == null || index >= s.length())
        return s;

    char toFlip = s.charAt(index);

    if(toFlip == '1')
        toFlip = '0';
    else
        toFlip = '1';
```

```
    StringBuilder sb = new StringBuilder(s);
    sb.setCharAt(index, toFlip);
    return sb.toString();
}
```

The following method is used to get a random binary string with a specific length.

```
public static String getRandomString(int length){

    if(length < 0)
        return null;

    String result = "";
    String now;

    for(int i = 0; i < length; i++){
        if(Math.random() > 0.49)
            now = "1";
        else
            now = "0";
        result+=now;
    }
    return result;
}
```

The next method is the method `split`; and we already encountered it in *section 1.3.1*, inside the method `Conversion`. Its aim is to split a string into several substrings, based on the indexes passed as parameter. It is slightly more complex than the other methods that we have just seen. In order to carry out its task, after some initial checks, the method has to sort the list of indexes. This operation is performed for us by the method `sort` of the interface `Comparator<T>`: all we need to do is to tell the comparator our sort criterion. After this initial task, the rest of the operations is pretty straightforward.

```
public static ArrayList<String> split(ArrayList<Integer> indexes,
                                                         String s){

    if(indexes == null || indexes.size() == 0 ||
                                         !indexes.contains(s.length())){
```

```
            return null;
        }

        //Defining the sort criterion and sorting
        indexes.sort(new Comparator<Integer>(){
            public int compare(Integer o1, Integer o2) {
            int a = o1, b = o2;
            if(a < b)
                return -1;
            else if (b < a)
                return 1;
            return 0;
            }
        });

        int lastIndex = indexes.get(indexes.size()-1);

        if(lastIndex > s.length())
            return null;

        Iterator<Integer> iterator = indexes.iterator();
        ArrayList<String> result = new ArrayList<String>();
        int prev = iterator.next();

        while(next.hasNext()){

            int current = iterator.next();
            String splitted = s.substring(prev, current);
            result.add(splitted);
            prev = current;
        }
        return result;
    }
```

As an example, we also present the method `generateKeys`. The aim of this method is to generate the keys $k_1^{new}$ and $k_2^{new}$, that are computed after a successful authentication of the Reader by the Tag. This method has been very useful when we had to test the effectiveness of the attack.

```
public static ArrayList<String> generateKeys(String k1, String k2,
                                       String n, String B, int treshold){
```

```
    String k1new = Utility.conversion(treshold, k1, n);
    k1new = Utility.stringXOR(k1new, k2);

    String k2new = Utility.conversion(treshold, k2, B);
    k2new = Utility.stringXOR(k2new, k1);

    ArrayList<String> keys = new ArrayList<>();
    keys.add(k1new);
    keys.add(k2new);

    return keys;
}
```

As the last method, we want to share an implementation of the algorithm to "invert" Conversion, that we presented in *section 2.4.1.1* . This implementation of the algorithm produces a $B$ string with three $1s$, so, it is not the easiest version of the procedure. We called the method `reverse` and it works with strings of 32 bits.

```
public static ArrayList<String> reverse(String desiredOutput, treshold){

    String white = "11100000000000000000000000000000";

    // this function calls Utility.flipBit on the bits at the positions
    // 0, 1 and 2 of the string desiredOutput to complement them.
    String secondString = flipAdjacentBits(0, 3, desiredOutput);

    // here we split the string secondString by the division's schema
    // generated by the string white.
    ArrayList<String> temp = Utility.split(
                Utility.computeDivision(white, treshold), secondString);
    secondString = "";

    // the following for cycle right-rotates every substring
    // contained in the array temp, by its Hamming-Weight.
    for(String t : temp)
        secondString += rightRotate(t, getHammingWeight(t));

    // we perform on the string white the operations that have already
    // been done on secondString
```

```
    temp = split(Utility.computeDivision(secondString, treshold), white);
    white = "";

    for(String t : temp)
        white += rightRotate(t, getHammingWeight(t));

    ArrayList inputs = new ArrayList<>();
    inputs.add(white);
    inputs.add(secondString);
    return inputs;
}
```